# Janus Monolayer Transition Metal Dichalcogenides

Jing Zhang[1‡], Shuai Jia[1‡], Kholmanov Iskandar[2], Liang Dong[3], Dequan Er[3], Weibing Chen[1], Hua Guo[1,] Zehua Jin[1], Vivek B. Shenoy[3], Li Shi[2] and Jun Lou[1*]

[1] Department of Materials Science and Nanoengineering, Rice University, Houston, TX, 77005

[2] Department of Mechanical Engineering, The University of Texas at Austin, Austin, TX, 78712

[3] Department of Materials Science and Engineering, University of Pennsylvania, Philadelphia, PA, 19104

[‡] The authors contribute equally to this work

Abstract: A novel crystal configuration of sandwiched S-Mo-Se structure (Janus SMoSe) at the monolayer limit has been synthesized and carefully characterized in this work. By controlled sulfurization of monolayer $MoSe_2$ the top layer of selenium atoms are substituted by sulfur atoms while the bottom selenium layer remains intact. The peculiar structure of this new material is systematically investigated by Raman, photoluminescence and X-ray photoelectron spectroscopy and confirmed by transmission-electron microscopy and time-of-flight secondary ion mass spectrometry. Density-functional theory calculations are performed to better understand the Raman vibration modes and electronic structures of the Janus SMoSe monolayer, which are found to correlate well with corresponding experimental results. Finally, high basal plane hydrogen evolution reaction (HER) activity is discovered for the Janus monolayer and DFT calculation implies that the activity originates from the synergistic effect of the intrinsic defects and structural strain inherent in the Janus structure.

Since the discovery of graphene in 2004[1], two-dimensional (2D) materials have been attracting increasing attention due to the many novel properties originating from the bulk to monolayer transition. Among the 2D family, transition metal dichalcogenides (TMDs) have been most widely studied; $MoS_2$ as the representative TMD material, its electrical[2], optical[3] as well as other physical properties[4,5] are well understood. By controlling the stoichiometric ratio of chemical vapor deposition grown $MoS_xSe_{2-x}$[6], $WS_xSe_{2-x}$[7] and $Mo_xW_{1-x}S_2$[8] alloys, the optical and electrical properties of the monolayers can be tuned. Various combinations of out-of-plane TMD heterojunctions have been prepared by transfer methods. $MoS_2/WS_2$[9], $MoS_2/MoSe_2$[10], $MoS_2/WSe_2$[11] and $MoSe_2/WSe_2$[12] out-of-plane and in-plane heterojunctions have also been grown by the CVD method. The heterojunctions, especially the in-plane ones, exhibit interesting physical properties due to the presence of an atomically sharp interface[13]. In this work we demonstrate a novel type of TMD structure possessing an unconventional asymmetry sandwich construction that is distinctively different from both types of heterojunctions. As shown in **Figure 1b** and **Figure 2a**, the so-called "Janus" SMoSe consists of three layers of atoms, namely, sulfur, molybdenum and selenium from the top to the bottom. Unlike its allotrope of the randomly alloyed SMoSe, the Janus SMoSe is highly asymmetric along the c-axis direction. The polarized chemical construction potentially derives an intrinsic electric field inside the crystal and novel physical properties such as the Zeeman-type spin splitting[14].

We have managed to reproducibly obtain monolayer Janus SMoSe flakes by well-controlled sulfurization of monolayer $MoSe_2$. The monolayer $MoSe_2$ flakes were first grown by the CVD method[15]. Briefly, $MoO_3$ powder as the Mo source was placed in a porcelain boat at the center of the heating zone. Sulfur powder as the S source was placed upstream of the heating zone. Monolayer $MoSe_2$ flakes started to crystallize and grow at around 800 °C on a $SiO_2$/Si substrate facing the $MoO_3$ powder under the protection of ultra-high purity argon. The as-grown monolayer $MoSe_2$ exhibits typical Raman and high photoluminescence[16,17] signals as shown in **Figure 1c** and **d**. The sulfurization of the top layer Se was realized by a controlled substitutional reaction with vaporized sulfur in a typical CVD setup (**Figure 1a**). Briefly, the monolayer $MoSe_2$ sample was placed in the center of the heating zone and heated to 800 °C. Sulfur powder was heated to 150 °C by a heating belt. The sulfur gas was carried to the center of the heating zone by ultra-high purity argon. The sulfurization process was kept for 30 min before the furnace was naturally cooled down to room temperature.

As shown in **Figure 1b**, the top and bottom layer Se atoms of monolayer $MoSe_2$ on $SiO_2$/Si substrate are located under different chemical environments, namely, under atmosphere pressure and in the Van der Waals gap, respectively. While the gaseous sulfur molecule (mainly $S_8$, possibly pyrolyse to smaller molecules) approaches, the top layer Se atoms are in direct contact with sulfur and could be quickly substituted once the thermal driving force overcomes the thermodynamic barrier. However, in order to substitute the bottom layer Se atoms, the gaseous sulfur molecules have to diffuse into the Van der Waals gap first. Once a selenium atom is substituted, it has to diffuse out to facilitate further sulfurization. Assuming the sulfur partial pressure is identical, the diffusion process becomes the rate-limiting step and the temperature directly determines the reaction rate. We have shown in **Figure 1b** that below 750 $^{o}$C and over 850 $^{o}$C the Raman peaks are similar to $MoSe_2$ and $MoS_2$ respectively, indicating that sulfur does not substitute selenium below a certain temperature and is able to diffuse into the Van der Waals gap once the temperature is high enough. Similar observation has been made on the sulfurization of monolayer $WSe_2$ as depicted in **Figure S2**. The tendency to form the Janus structure has been shown in similar sulfurization process but under high vacuum condition[7,18,19]. The high vacuum potentially lowers the diffusion energy barrier of sulfur and facilitates the sulfurization process, resulting in a reduced temperature required for complete sulfurization of the selenides (~700 $^{o}$C). We attribute our reproducible synthesis of the Janus SMoSe to the fact that atmosphere pressure is maintained kept during the reaction. The atmospheric pressure significantly enlarges the stable temperature window for the top layer atom substitution reaction. The sulfurization time, on the other hand, did not lead to the selenium substitution in the bottom layer, implying that only temperature and pressure played crucial roles in this selective sulfurization process.

The Janus SMoSe triangular flake looks identical to $MoSe_2$ when observed under the optical microscope (**Figure 2b**). The morphology of the flake was checked by atomic force microscopy (AFM) and the profile shows the flake thickness was less than 1nm (**Figure 2e**), implying after sulfurization it is still in the monolayer configuration. To interpret the tri-layer-atom structure we first examined the Raman spectra of the Janus SMoSe flakes. As shown in **Figure 2a**, the spectrum for the as-grown $MoSe_2$ has the typical $A_{1g}$ mode (240 cm$^{-1}$) and $E^1_{2g}$ mode (287 cm$^{-1}$) peaks[17,20]. After the sulfurization reaction, several major changes appeared at the spectrum. First of all, the $A_{1g}$ peak completely vanished, indicating the out-of-plane vibration of the symmetric Se-Mo-Se structure has been disrupted by the sulfur substitution to selenium. Secondly, a new peak with strong intensity appeared at 290 cm$^{-1}$, not overlapping with the $MoSe_2$ $E^1_{2g}$ peak (287 cm$^{-1}$). We suspect the peak accounts for the out-of-plane S-Mo-Se vibration mode. At the same time, another distinct peak that we assign for the in-plane S-Mo-Se vibration mode at 350.8 cm$^{-1}$ appeared. In the end, compared to the $MoS_xSe_{2-x}$ alloy spectrum, which consists of the $MoS_2$ and $MoSe_2$ major Raman peaks as long as $0<x<1$[6,18], none of the representative peaks for $MoS_2$ nor $MoSe_2$ could be observed. The absence of those peaks clearly indicates that the randomly distributed alloy structure does not exist in our sample. The Raman spectrum for the sulfurized sample is consistent with our expectation of the Janus SMoSe structure. Raman peak intensity mapping in **Figure 2c** at the main vibration peak for Janus SMoSe (290 cm$^{-1}$) clearly demonstrates that the flake is uniformly transformed from $MoSe_2$ to Janus SMoSe. We have also shown the layer number dependence of the Raman signal of $MoSe_2$ multilayers (**Figure S3**). Once the layer number exceeds two, the Raman signal is a mixture of Janus SMoSe and $MoSe_2$. The result clearly indicates that at 800 °C sulfur is able to substitute the surface selenium solely and the atom layers are arranged following the S-Mo-Se Se-Mo-Se… sequence in the final product.

The existence of sulfur in the Janus SMoSe was confirmed by X-ray photoelectron spectroscopy. Survey scan (**Figure S1**) reveals the existence of S, Se and Mo in the material and Si, O from the substrate. The S 2p1/2 and 2p3/2 peaks fitted from the XPS results (**Figure 2j**) are located at 161.3 eV and 162.5 eV, in accordance with the S peak position for sulfides[21]. The Se 3d3/2 and 3d5/2 peaks (**Figure 2i)** are located at 54.6 eV and 53.8 eV matching the selenides peak position range as well. The two extra peaks (**Figure 2j**) are assigned to Se 3p1/2 (166.1 eV) and 3p3/2 (160.2 eV). The Mo 3d3/2 and 3d5/2 peaks (**Figure 2h**) are located at 231.6 eV and 228.5 eV in the $MoS_2$ peak positon range. The atomic ratio of S, Mo and Se was estimated to be 1.02:1.05:1 according to the peak areas of the S 2p, Mo 3d and Se 3d peaks, which is in good agreement with the stoichiometric ratio (1:1:1) of those elements in the Janus SMoSe crystal. The Janus SMoSe remained in 2H crystal structure, according to the diffraction pattern (**Figure 2g**) and HRTEM image (**Figure 2f**) of the monolayer, indicating there is neither significant lattice distortion nor phase transition induced by the top layer sulfur atom substitution of selenium.

Density functional theory (DFT) simulations were carried out to study the lattice vibration and electronic band structure of the Janus SMoSe. As shown in **Figure 3a**, the calculated $A_{1g}$ and $E_{2g}^1$ modes of the Raman spectrum have the frequencies of 234.9 cm$^{-1}$ and 276.9 cm$^{-1}$, respectively, for $MoSe_2$ monolayer. The calculated $A_{1g}$ and $E_{2g}^1$ modes for the Janus SMoSe monolayer are at 283.1 cm$^{-1}$ and 343.5 cm$^{-1}$, respectively. The calculated $A_{1g}$ and $E_{2g}^1$ modes of the Raman spectrum have the frequencies of 396.3 cm$^{-1}$ and 373.1 cm$^{-1}$, respectively, for $MoS_2$ monolayer. As listed in **Table S1**, the experimental and computational results are in perfect agreement, confirming the origin of the characteristic peaks from the Raman spectra and our interpretation of the Janus SMoSe structure. Moreover, a strong activity for SMoSe at $B_{2g}^1$ (**Figure S4**) 427.8 cm$^{-1}$ is predicted according to the simulation. This mode is inactive for $MoS_2$ and $MoSe_2$, but is active for SMoSe because of the broken symmetry along the z-direction. The peak could be observed at 436.4 cm$^{-1}$ in our experimental spectra (**Figure 1c**). This peak becomes noticeable while the starting material before sulfurization is double and multilayer $MoSe_2$ (**Figure S3b**), implying the peak is significantly affected by the phonon scattering originating from the material/substrate interaction[22].

As shown in **Figure 3b** and **Figure S5**, the Janus SMoSe is an indirect band gap semiconductor, with the conduction band minimum located at the K point in the reciprocal space, while the valence band minimum is at the Γ point. This is distinctively different from either monolayer $MoS_2$ or $MoSe_2$ that has a direct band gap (K point – K point). The calculated K point – K point transition energy (1.568 eV) in the Janus SMoSe is slightly larger than its indirect band gap energy (1.558 eV, Γ point – K point), implying that the radiative electron-hole recombination in this material in the PL experiment can still be observed, but with significantly reduced probabilities. We inspected the photoluminescence spectra of the pure $MoSe_2$ and Janus SMoSe samples and the results are in good agreement with the theoretical predictions. In **Figure 1c** the peak position of Janus SMoSe not only blue-shifted but also quenched, implying the larger bandgap energy of the structure comparing to pure $MoSe_2$ as well as the direct to indirect semiconductor transition. The PL intensity mapping of the Janus SMoSe triangular flake at 1.68 eV is shown in **Figure 2d**. Similar to the Raman intensity mapping, the PL intensity mapping displays a uniform distribution, indicating the high level of homogeneity of the material.

Time-of-flight secondary ion mass spectrometry (TOF-SIMS) can reveal the elemental distribution along the basal plane direction of 2D materials as well as depth profile with atomic resolution. The SIMS images showed the uniform distributions of S, Mo and Se ions based on the overall signal collected for each element (**Figure 4a-c**). The depth profile of the Janus SMoSe sample displays clearly the knockout sequence of S, Mo, Se and $SiO_2$ from the top to the bottom of the flake. In contrast, the pure $MoSe_2$ sample displays a distinguishing knockout sequence (**Figure S6**). Therefore, we prove that the Janus SMoSe is composed of S, Mo and Se atom layers as we inferred and exclude the possibility that the Janus SMoSe resemble the $MoS_xSe_{2-x}$ alloy structure.

Group IV transition metal dichalcogenides are known for their promising catalytic activities, particularly for the hydrogen evolution reaction (HER). The edge sites contribute to major catalytic activity while the basal plane sites are inactive. Here we demonstrated that the basal plane HER activities of the Janus SMoSe and its reverse configuration SeMoS are stimulated by the controlled atom substitution comparing with pure $MoS_2$ and $MoSe_2$. (**Figure 5a**). The Janus layers of SMoSe and its reverse construction SeMoS both exhibit lower basal plane HER overpotentials than $MoS_2$ and $MoSe_2$. Meanwhile, the performance of SeMoS basal plane surpasses that of SMoSe. The calculated results of hydrogen adsorption free energy $\Delta G_H$ are reported in **Figure 5b**, where the highest value (0.161 eV) occur in the $MoS_2$ supercell (with a single S-vacancy), while the lowest value (-0.007 eV, very close to charge neutrality condition) is obtained in the SeMoS supercell (with a single Se-vacancy). Values of $\Delta G_H$ for SMoSe (with a single S-vacancy) and for $MoSe_2$ (with a single Se-vacancy) are comparable (0.060 eV and 0.063 eV, respectively). These values are significantly lower than $\Delta G_H \approx 2$ eV on the surfaces of pristine $MoS_2$ supercell without any vacancies[23], implying that the presence of single S- and Se-vacancies significantly strengthen the binding of H with the TMD monolayers and hence turn the inert surfaces to be catalytically active for HER. According to our results, the HER efficiencies of these TMDs should be SeMoS>SMoSe≈$MoSe_2$>$MoS_2$, in general agreement with the experimental results that SeMoS>SMoSe>$MoSe_2$>$MoS_2$. The only exception is the relative efficiencies of SMoSe and $MoSe_2$. Theoretical results show that they have similar HER efficiencies while experimentally SMoSe has a better efficiency. The discrepancy may arise from the densities of single S- and Se-vacancies, which affect the electronic and catalytic properties of TMD monolayers. In the supercells we simulate, a universal vacancy density of 6.25% in an S or Se atomic layer is used, regardless of the materials. However, under experimental conditions, the single S- and Se-vacancies may vary in $MoS_2$, SMoSe (SeMoS), and $MoSe_2$. Therefore, a further experimental investigation on the defect microstructure of the TMD samples is needed to address this discrepancy.

**Conclusion**
2D monolayer with novel asymmetric structure – Janus SMoSe has been synthesized by a controlled sulfurization process. The S-Mo-Se tri-layer atomic structure exhibits unique Raman vibration peaks that can be distinguished from pure $MoS_2$, $MoSe_2$ and randomly alloyed $MoS_xSe_{2-x}$. The experimental characteristic Raman peaks agree well with predictions of DFT simulations. TOF-SIMS analysis further confirms that the Janus

SMoSe is composed of the S-Mo-Se tri-layer structure. Both Raman mapping and TOF-SIMS element mapping indicate that the monolayer is uniformly sulfurized and the asymmetric structure distributes all over the Janus monolayer. DFT simulation was used to predict the bandgap of the Janus SMoSe and the value is consistent with the experimental results derived from the PL spectroscopy. It is also suggested that an intrinsic electric field potentially exists perpendicular to the inplanar direction. High basal plane HER catalytic activity is discovered for the Janus structures and DFT calculation implies that the activity may originate from the synergistic effects of the intrinsic defects and structural strain in the Janus structure.

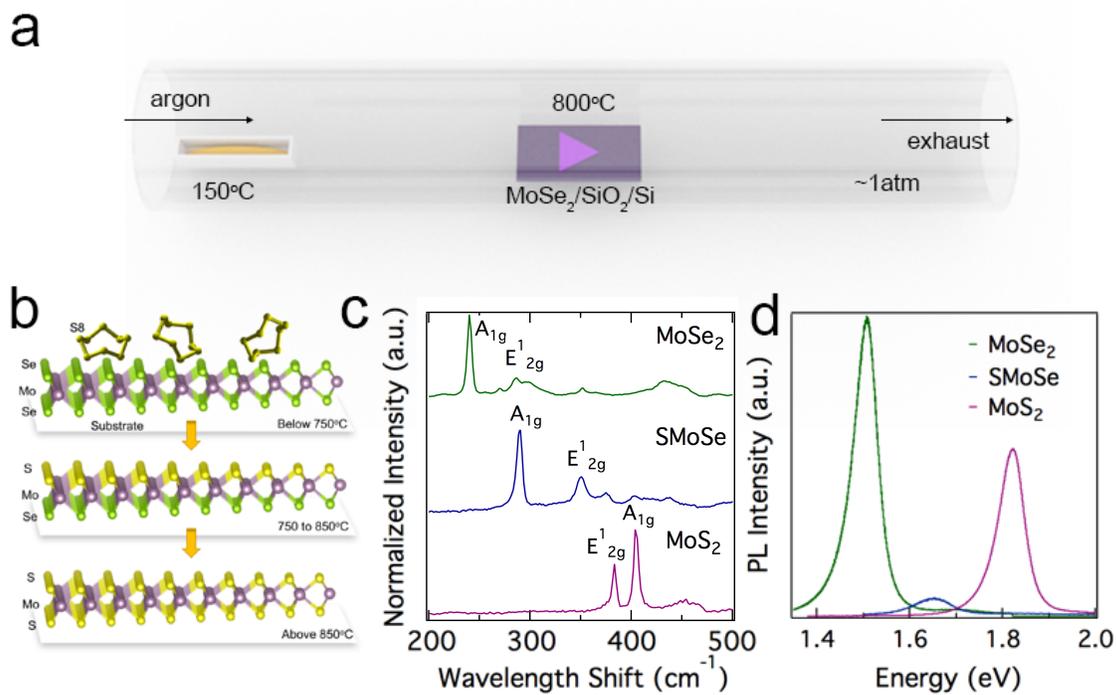

**Figure 1**. (a) Schematic illustration of the reaction setup. (b) Proposed reaction mechanism for the sulfurization of monolayer $MoSe_2$ on a $SiO_2/Si$ substrate at different temperatures. Between the monolayer and the substrate is the Van der Waal interaction. (c)&(d) Raman and PL (under 532 nm diode laser excitation) spectra of $MoSe_2$, Janus SMoSe and $MoS_2$ corresponding to the diagram in (a).

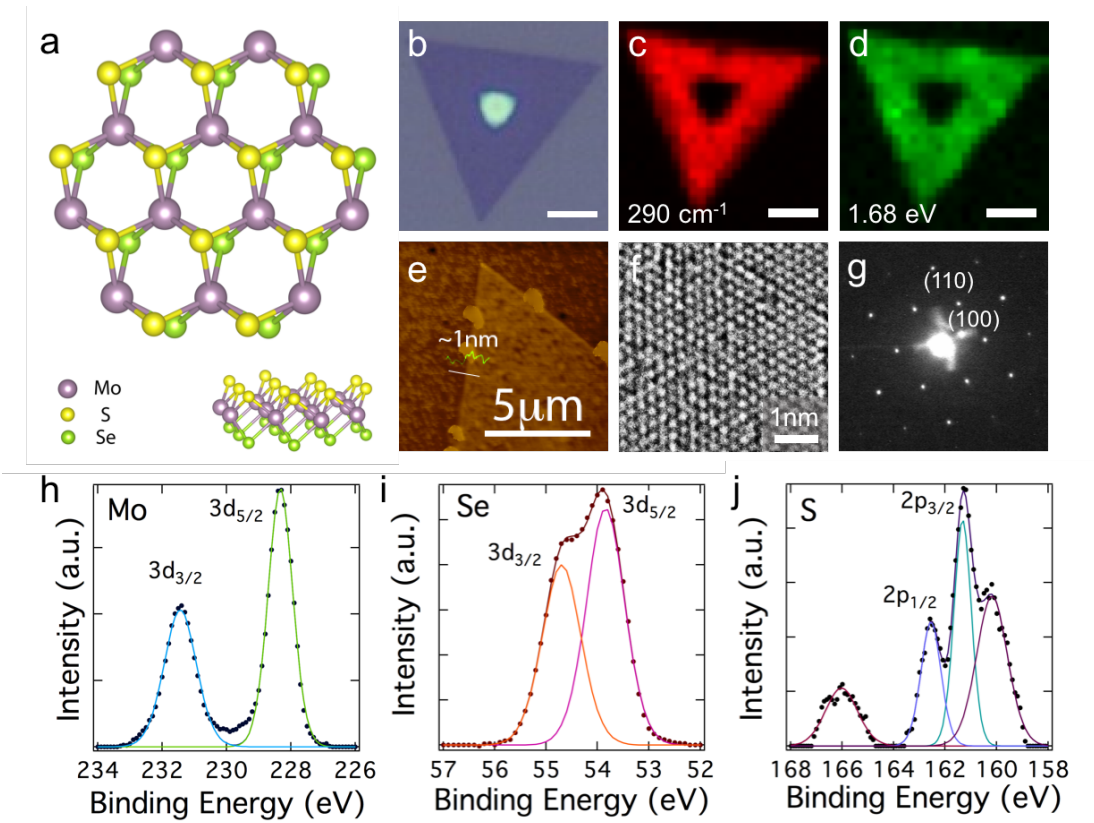

**Figure 2**. (a) Off angle top view and side view of an eight-unit-cell Janus SMoSe monolayer. The purple, yellow and green spheres represent molybdenum, sulfur and selenium atoms respectively. (b) Optical image of a Janus SMoSe triangle. The purple and the central island with high contrast is the monolayer and bulk crystal region respectively. (c)(d) Raman and PL peak intensity mappings of the Janus SMoSe triangle in (b). The mapping shows uniform distribution of the identical Raman peak at 287 cm$^{-1}$ and PL peak at 1.68 V. (e) AFM topography image of the Janus SMoSe triangle. The profile shows that the thickness of the flake is less than one nanometer. (f) HRTEM image of the Janus SMoSe lattice. The atom arrangement indicates the 2H structure of the monolayer. (g) Corresponding SAED pattern of the monolayer. (h-j) XPS spectra of the Mo 3d, Se 3d and S 2p core level peaks for the Janus SMoSe monolayer.

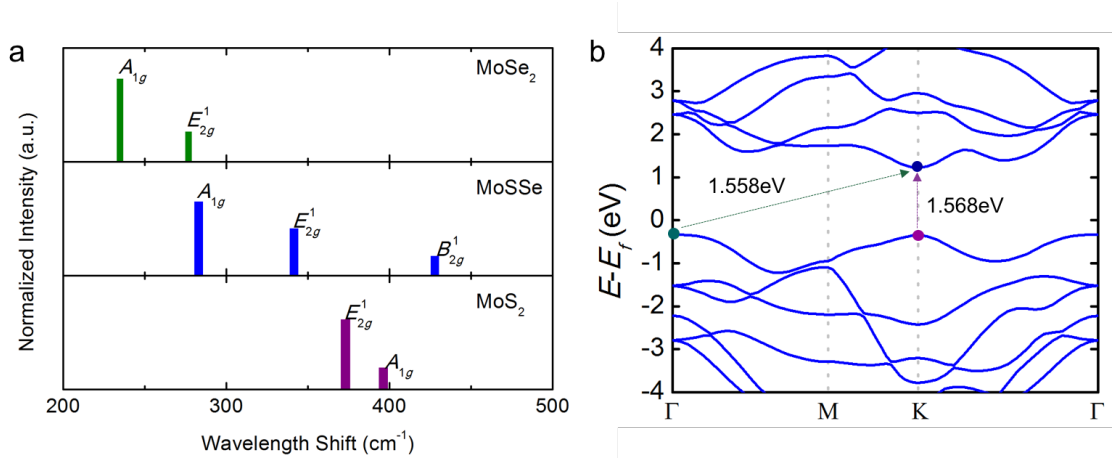

**Figure 3**. Computational evaluation on the Janus SMoSe structure. (a) Predicted Raman vibration modes on the monolayer MoSe$_2$, Janus SMoSe and MoS$_2$. Three major vibration modes A$_{1g}$, E$^1_{2g}$ and B$^1_{2g}$ are proposed for the Janus SMoSe, locating at 283.1 cm$^{-1}$, 341.5 cm$^{-1}$ and 427.8 cm$^{-1}$, respectively. (b) Predicted band diagram of monolayer Janus SMoSe. The direct and indirect bandgaps at K and Γ points are calculated to be 1.568 eV and 1.558 eV, respectively.

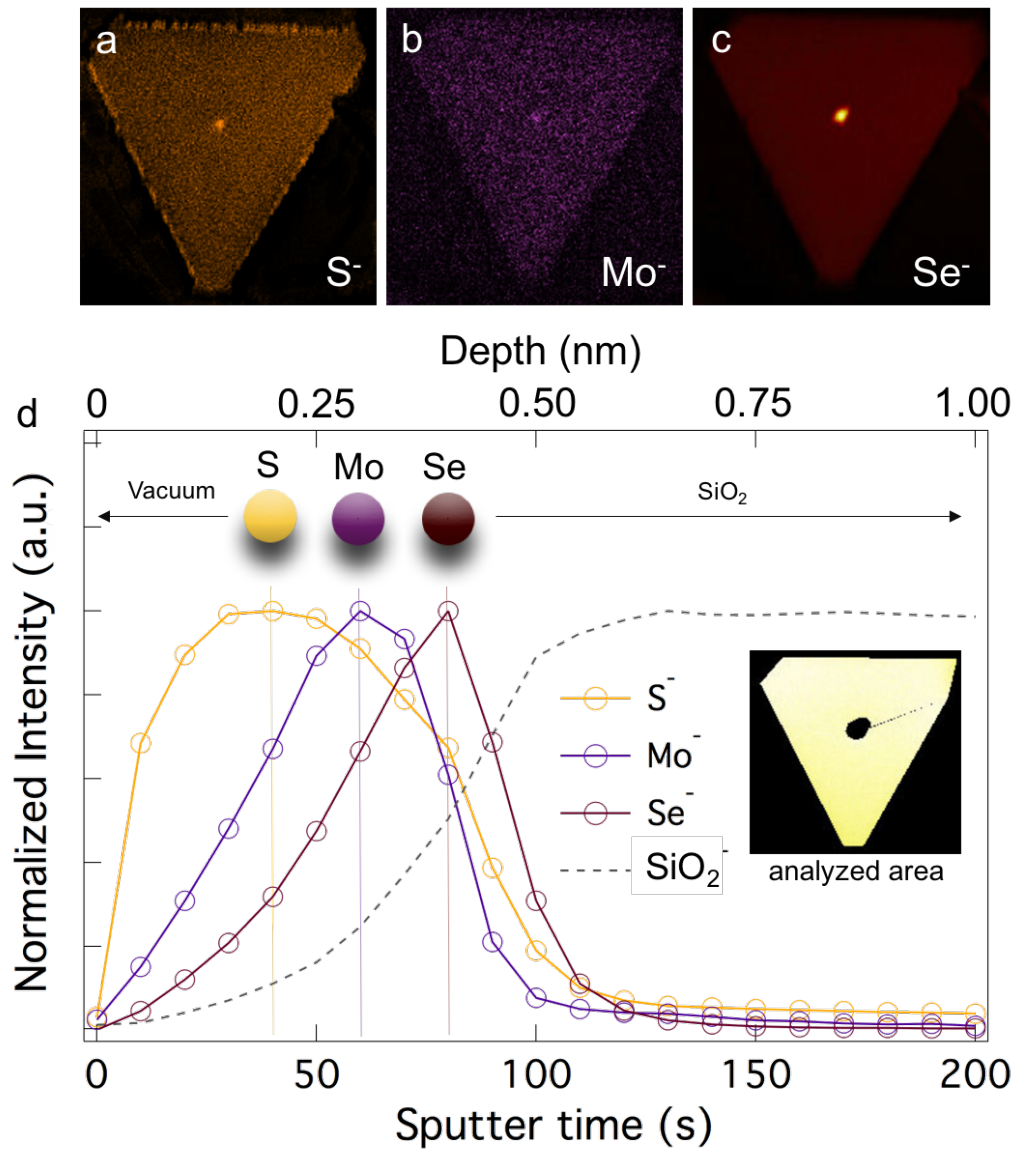

**Figure 4**. TOF-SIMS analysis of the Janus SMoSe triangular flake. (a)-(c) SIMS images of the sulfur, molybdenum and selenium ions from a typical SMoSe triangular flake. Brighter color represents high concentration of atoms. (d) Normalized TOF-SIMS depth profile of sulfur, molybdenum and selenium ions of the SMoSe triangular flake. The analyzed area is shown in the inset.

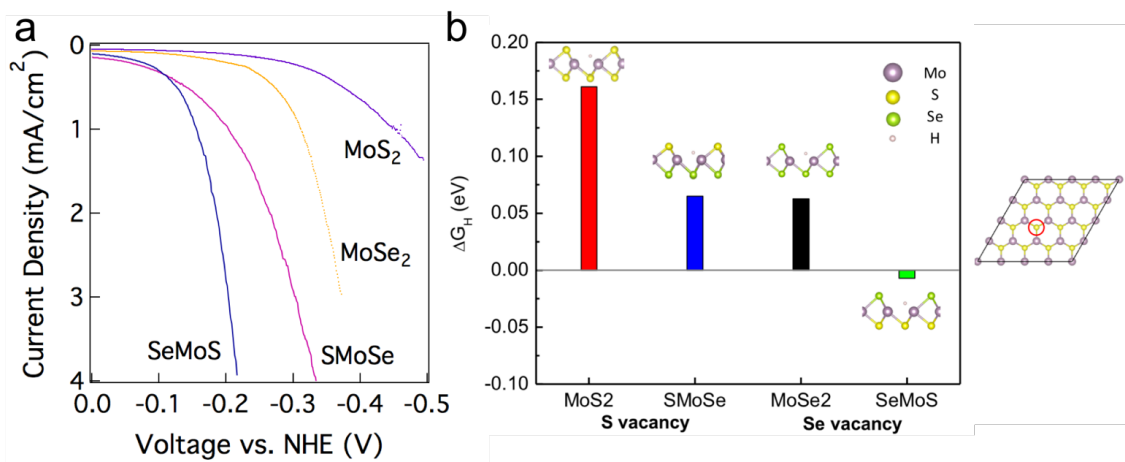

**Figure 5**. (a) Hydrogen evolution reaction polarization curves of monolayer MoS2, MoSe2, SMoSe and SeMoS. (b) (Left) the DFT calculated hydrogen adsorption free energy $\Delta G_H$ for $MoS_2$ and SMoSe (which have a single S-vacancy), and for $MoSe_2$ and SeMoS (which have a single Se-vacancy). The insets show a side view of the truncated supercell structure where the H atom is absorbed in the vacancies. (Right) the top view of the 4×4×1 supercell with the location of the single S- (or Se-) vacancy circled in red.

# Janus Monolayer Transition Metal Dichalcogenides

Jing Zhang[1‡], Shuai Jia[1‡], Kholmanov Iskandar[2], Liang Dong[3], Dequan Er[3], Weibing Chen[1], Hua Guo[1], Zehua Jin[1], Vivek B. Shenoy[3], Li Shi[2] and Jun Lou[1*]

**Experimental Details**
All the chemicals are purchased from Sigma-Aldrich and used as received without further purification. The CVD procedures are carried out using a MTI tube furnace. The Raman and PL spectra are collected with Renishaw inVia confocal Raman microscope. The TEM images are taken with JOEL 2100F microscope. The XPS spectra are collected with PHI Quantera instrument. The devices are fabricated with FEI Quanta 400 SEM. The TOF-SIMS data is collected with a IONTOF TOF.SIMS.5 instrument.

The electronic properties of such asymmetric 2D monolayer is investigated by measuring the field-effect transistor (FET) based on it. FET devices on Janus SMoSe monolayers were fabricated using e-beam lithography method. The fabricated device (inset in **Figure S6**) has a channel length of 6 μm and width of 3.6 μm. At positive back-gate voltage ($V_g$) region the output characteristics (increased drain current ($I_d$) vs. $V_g$) indicates the n-type behavior of the Janus SMoSe monolayer. The mobility of the device is calculated (see details in supplementary information) to be $6.2 \times 10^{-2}$ cm$^{-2}$ V$^{-1}$ s$^{-1}$ and the on/off ratio is $10^2$. The low mobility of the Janus SMoSe can potentially be attributed to the heat treatment accompanying the sulfurization reaction. During the process the residual oxygen in the CVD environment often lead to the creation of vacancies on the monolayer structure[1,2] and reduction in the carrier mobility[3].

The basal plane HER polarization curves of the monolayer TMDs were collected using a modified three-electrode method. The TMD monolayers were electrically connected by depositing gold electrodes on them. To exclude the edge sites contribution of the monolayers to the HER measurements, a layer of PMMA was coated onto a typical HER device and patterned with e-beam lithography method so that only a basal plane region with controlled area was exposed for the HER measurement.

**Descriptor of the HER Efficiency**
The catalytic efficiency of TMDs for HER is characterized by hydrogen adsorption free energy $\Delta G_H$[4]. A positive $\Delta G_H$ indicates weak absorption of hydrogen atoms on the surfaces of TMDs in the initial stage of HER, which reduces the probability of subsequent reactions. On the contrary, a negative $\Delta G_H$ indicates strong binding of H atoms on the surfaces, which prevents the detachment of H$_2$ molecules from the catalyst. The optimal value of $\Delta G_H$ is 0 eV, i.e., thermally neutral. $\Delta G_H$ is defined as

$$\Delta G_H = \Delta E_H + \Delta E_{ZPE} - T\Delta S. \qquad (1)$$

In the above equation, $\Delta E_H$ is the differential hydrogen adsorption energy, and is calculated by

$$\Delta E_H = E(SMoSe + H) - E(SMoSe) - \frac{1}{2}E(H_2), \qquad (2)$$

where $E(SMoSe + H)$, $E(SMoSe)$, and $E(H_2)$ represent the total energy of SMoSe with one H atom absorbed on the surface, the energy of SMoSe with clean surface, and the energy of a H$_2$ molecule, respectively. $\Delta E_{ZPE}$ in Eq. (1) is the difference in zero point energy between the adsorbed hydrogen and hydrogen in the gas phase, and is calculated from the vibrational frequencies of the H atom at 0K. Finally, T and ΔS in Eq. (1) are the temperature (set to be 300K) and the entropy difference between absorbed H and H in the gas phase, respectively. In our calculations, the configurational entropy of the H atom in the adsorbed state is small and is neglected. Therefore, $\Delta S = -1/2\, S(H_2)$ where $S(H_2)=130.68$ J·mol$^{-1}$·K$^{-1}$ is the entropy of the $H_2$ molecule in the gas phase at standard conditions: 1 bar of $H_2$, pH = 0 and T = 300 K[5].

**Computational Details**

The DFT calculations are performed using the Vienna ab-initio simulation package (VASP) code[6]. Projector augmented wave pseudopotentials[7] are used with a cutoff energy of 520 eV for plane-wave expansions. The exchange-correlation functional was treated within the Perdew–Burke–Ernzerhof (PBE) generalized gradient approximations (GGA)[8]. The unit cell structures of monolayer SMoSe, $MoS_2$ and $MoSe_2$ are relaxed with a Γ-centered *k*-point mesh of 18×18×1 in the first Brillouin zone, with the total energy converged to below $10^{-8}$ eV. The in-plane lattice parameters of $MoSe_2$, SMoSe, and $MoS_2$ monolayers are calculated to be 3.322 Å, 3.252 Å, and 3.185 Å, respectively. The atomic positions of the unit cells are optimized until all components of the forces on each atom are reduced to values below 0.001 eV/Å. A vacuum region of 16 Å thickness was used to prevent interactions between the adjacent periodic images of the monolayer. To simulate the Raman activity of these materials, their dielectric susceptibility tensor and zone-centered (Γ point) vibrational frequencies are computed with the density functional perturbation theory as implanted in VASP. Raman intensities are estimated by the derivative of the dielectric susceptibility tensor with respect to the normal mode[9], using the using the vasp_raman.py package[10].

To correctly address the chemisorption energies of H atoms on the surfaces of SMoSe, we use the Bayesian error estimation exchange-correlation functional with van der Waals interactions (BEEF-vdW)[11] as implemented in VASP with spin-polarized calculations. Other sets of parameters remain the same as in previous calculations for the structural relaxations. The in-plane lattice parameters of $MoSe_2$, SMoSe, and $MoS_2$ monolayers obtained by the BEEF-vdw functional are 3.335 Å, 3.263 Å, and 3.185 Å, respectively. 4×4×1 supercells of these materials (see right panel of Fig. X) are constructed with a single S- or Se- vacancy to investigate their effect on the HER catalytic efficiency. Such vacancies are most energetically favored among all types of structural defects in TMDs, and are hence mostly commonly observed in CVD-grown TMD sample[3]. Our calculations show that the single S- and Se-vacancies are the most energetically stable H absorption site on the surfaces of the supercells (see insets of the left panel of Fig. X). The vibrational frequencies of the absorbed H atom are calculated using the finite displacement method. The calculated $\Delta E_{ZPE}$ for H in a single S- and Se-vacancies in $MoSe_2$, SMoSe, and $MoS_2$ are 5.6 meV and -0.8 meV, respectively.

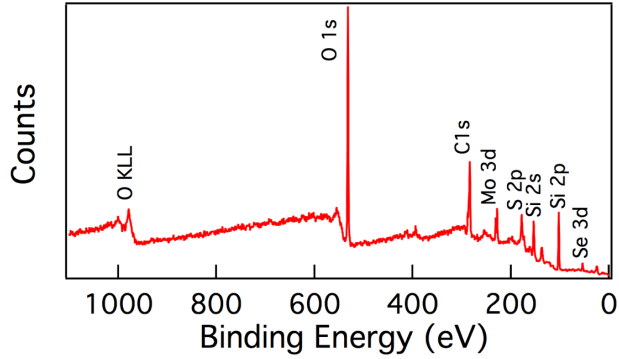

**Figure S1.** Survey scan of the Janus SMoSe monolayer on SiO$_2$/Si wafer.

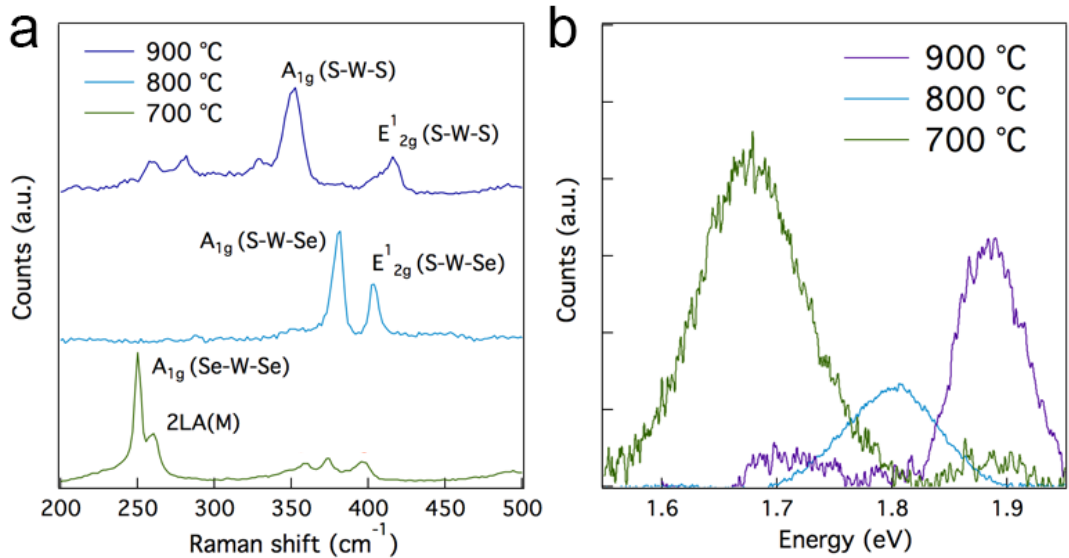

**Figure S2.** (a) Raman and (b) PL spectra of WSe$_2$ after being sulfurized at different temperatures. At 700 °C (green) the Raman spectra remains similar to pristine WSe$_2$, with emerging peaks originated from the scattered surface selenium substitution located between 350 to 400 cm$^{-1}$. At 800 °C (blue) two characteristic peaks are observed, indicating the formation of the Janus SWSe structure. At 900 °C (purple) the spectra evolves into the WS$_2$ analogue, implying the complete selenium substitution by sulfur. The PL spectra show the similar evolution trend with MoSe$_2$.

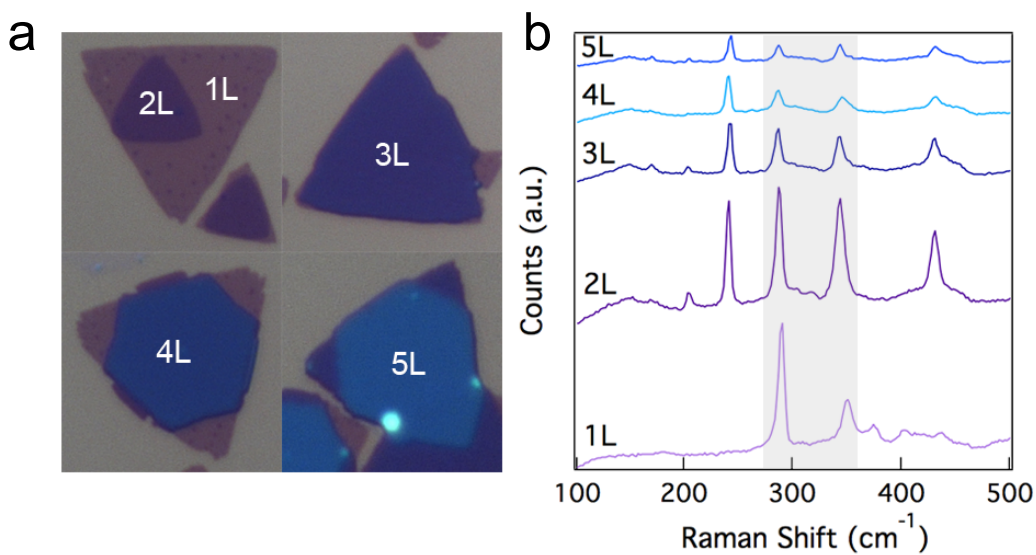

**Figure S3**. (a) MoSe$_2$ flakes with increasing number of layers. The contract of the MoSe$_2$ develops from light purple to light blue as the number of layer increases. (b) Layer number dependence on the Raman of Janus SMoSe on MoSe$_2$ flakes. The gray background region highlights the evaluation of the Raman characteristic peaks of the Janus SMoSe.

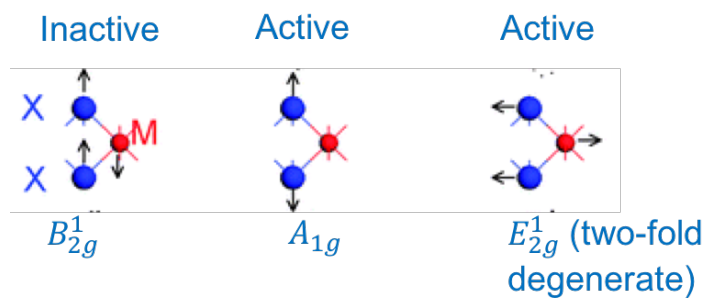

**Figure S4**. Schematic illustration of the lattice vibration modes in MX$_2$ metal dichalcogenides. The B$^1_{2g}$ mode is only active while the top and bottom atoms are not identical.

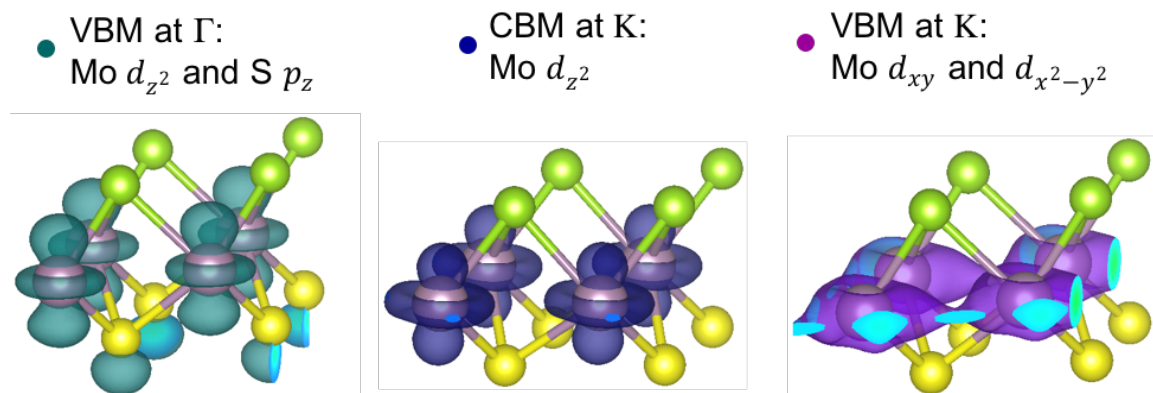

**Figure S5**. Charge density distribution of VBM at Γ point (green), CBM at K point (blue) and VBM at K point (purple) of monolayer Janus SMoSe.

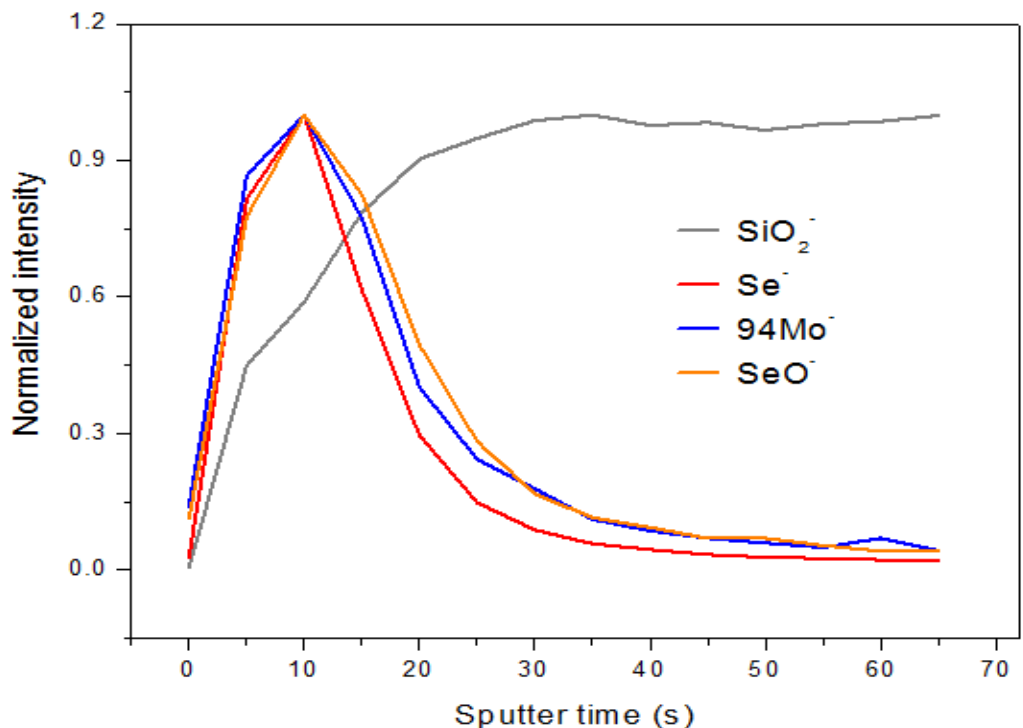

**Figure S6**. TOF-SIMS of pure MoSe$_2$.

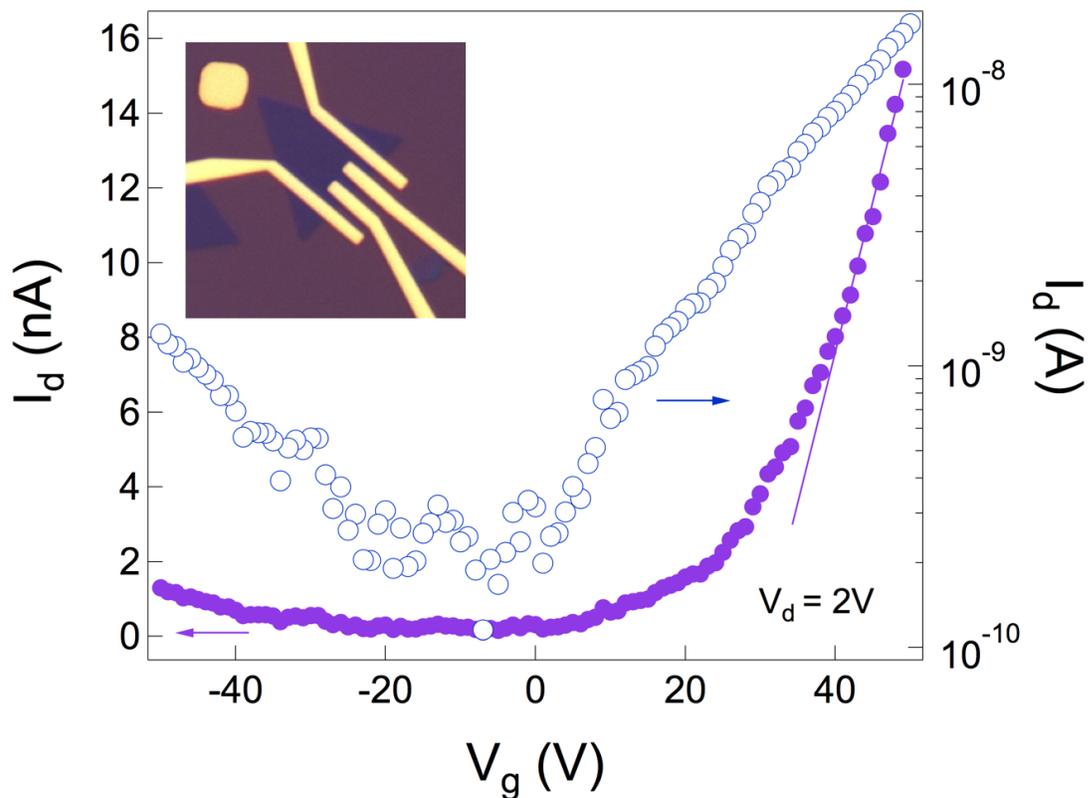

**Figure S7**. (Solid circles) Drain-source current Id as a function of back-gate voltage $V_g$ at fixed drain-source bias voltage $V_d$ = 2 V. (Solid line) Linear fit of the data within the back gate voltage range from 40 – 50 V. The carrier mobility based on the linear fit data is calculated to be m = 0.06

cm$^2$ V$^{-1}$ s$^{-1}$. (Hollow circles) Corresponding I$_d$ plotted in logarithmic scale as a function of back-gate voltage V$_g$ at fixed drain-source bias voltage V$_d$ = 2 V. Inset shows the optical image of a typical Janus SMoSe field-effect transistor device.

**Table S1**. Summary of the experimental and computational results regarding on the A$_{1g}$ and E$^1_{2g}$ vibrations modes of the monolayer MoSe$_2$, Janus SMoSe and MoS$_2$.

|  | Experiment (cm$^{-1}$) | | Computation (cm$^{-1}$) | |
| --- | --- | --- | --- | --- |
|  | $A_{1g}$ | $E^1_{2g}$ | $A_{1g}$ | $E^1_{2g}$ |
| MoSe$_2$ | 240 | 287 | 234.9 | 276.9 |
| SMoSe | 290 | 350.8 | 283.1 | 341.5 |
| MoS$_2$ | 404 | 383 | 396.3 | 373.1 |